\documentclass[12pt,preprint]{aastex}

\begin{document}

\title{Variation of Galactic Bar Length with Amplitude and Density
as Evidence for Bar Growth over a Hubble Time}

\author{Bruce G. Elmegreen}\affil{IBM Research Division, T.J. Watson
Research Center, 1101 Kitchawan Road, Yorktown Heights, NY 10598,
USA; bge@watson.ibm.com}

\author{Debra Meloy Elmegreen}\affil{Vassar College, Dept. of
Physics \& Astronomy, Box 745, Poughkeepsie, NY 12604, USA;
elmegreen@vassar.edu}

\author{Johan H. Knapen}\affil{Instituto de Astrof\'\i sica de
Canarias, E-38200 La Laguna, Spain; jhk@iac.es}

\author{Ronald J. Buta}\affil{Department of Physics and Astronomy,
University of Alabama, Tuscaloosa, AL 35487, USA;
buta@sarah.astr.ua.edu}

\author{David L. Block}\affil{School of Computational \& Applied Mathematics
University of the Witwatersrand P.O Box 60 Wits, 2050, South Africa;
block@wits.cam.ac.za}

\author{Iv\^anio Puerari}\affil{Instituto Nacional de Astrof\'{\i}sica,
Optica y Electr\'onica, Tonantzintla, PUE 72840, Mexico;
puerari@inaoep.mx}

\begin{abstract}
$K_s$-band images of 20 barred galaxies show an increase in the peak
amplitude of the normalized $m=2$ Fourier component with the
$R_{25}$-normalized radius at this peak.  This implies that longer
bars have higher $m=2$ amplitudes. The long bars also correlate with
an increased density in the central parts of the disks, as measured
by the luminosity inside $0.25R_{25}$ divided by the cube of this
radius in kpc. Because denser galaxies evolve faster, these
correlations suggest that bars grow in length and amplitude over a
Hubble time with the fastest evolution occurring in the densest
galaxies. All but three of the sample have early-type flat bars;
there is no clear correlation between the correlated quantities and
the Hubble type.
\end{abstract}

\keywords{galaxies: structure --- galaxies: spiral}

\section{Introduction} \label{sect:intro}

Bars should slow down and grow over time as bar angular momentum is
transferred to the disk (Tremaine \& Weinberg 1984) and halo
(Kormendy 1979; Sellwood 1980; Little \& Carlberg 1991; Hernquist \&
Weinberg 1992; Debattista \& Sellwood 1998, 2000; Valenzuela \&
Klypin 2003; Athanassoula 2002, 2003). With this growth, the bars
should become stronger, longer and thinner (Athanassoula 2003).

Pattern speeds are difficult to measure (Knapen 1999) but bar
lengths are not (Erwin 2005). To investigate the model predictions,
we examined relative bar lengths and intensities in 20 galaxies with
conspicuous bars and a range of Hubble types. We consider how these
parameters correlate with each other and with the central density of
the galaxy.  Central luminosity density is used as an indirect
measure of the inner angular rotation rate because few galaxies in
our sample have observed rotation curves. Galaxies with high central
densities should have high central rotation rates and evolve more
quickly than galaxies with low central densities. If there is a
secular change in bar length or amplitude with angular momentum
transfer, then denser galaxies should show the later evolutionary
stages.

\section{Observations and Analysis}

$K_{\rm s}$-band images of barred galaxies were obtained with the
Anglo-Australian Telescope (AAT) from 2004 June 28 to July 5. We
used the Infrared Imager and Spectrograph (IRIS2) with a $1024\times
1024$ pixel Rockwell HAWAII-1 HgCdTe detector mounted at the AAT's
$f/8$ Cassegrain focus, yielding a pixel scale of
0.447\,arcsec\,px$^{-1}$ and a field of view of 7.7\,arcmin squared.
Exposure times were around one hour in almost all cases and the
angular resolution was typically 1.5\,arcsec. Full details of the
observations will be presented in Buta et al. (2007).

Images were pre-processed using standard IRAF\footnote{IRAF is
distributed by the National Optical Astronomy Observatory, which is
operated by the Association of Universities for Research in
Astronomy, Inc., under cooperative agreement with the National
Science Foundation. } routines, and each image was cleaned of
foreground stars and background galaxies. Deprojections were derived
as follows. For each galaxy, estimates of the orientation parameters
were obtained using an ellipse fitting routine, {\it sprite},
originally written by W. D. Pence.
These fits were either based on the $K_s$-band image itself, or on
an optical image if available. Because the bulges may not be as flat
as the disks, we used a two-dimensional multi-component
decomposition code (Laurikainen, Salo, \& Buta 2005) to derive the
parameters of the bulges and disks. Images were deprojected,
assuming the bulges are spherical, using the IRAF routine IMLINTRAN.
This assumption has little impact on our Fourier analyses. The
results of the decompositions, as well as the orientation parameters
used, will be presented in Buta et al. (2007).

\section{Results}

\begin{figure}\epsscale{1}
\plotone{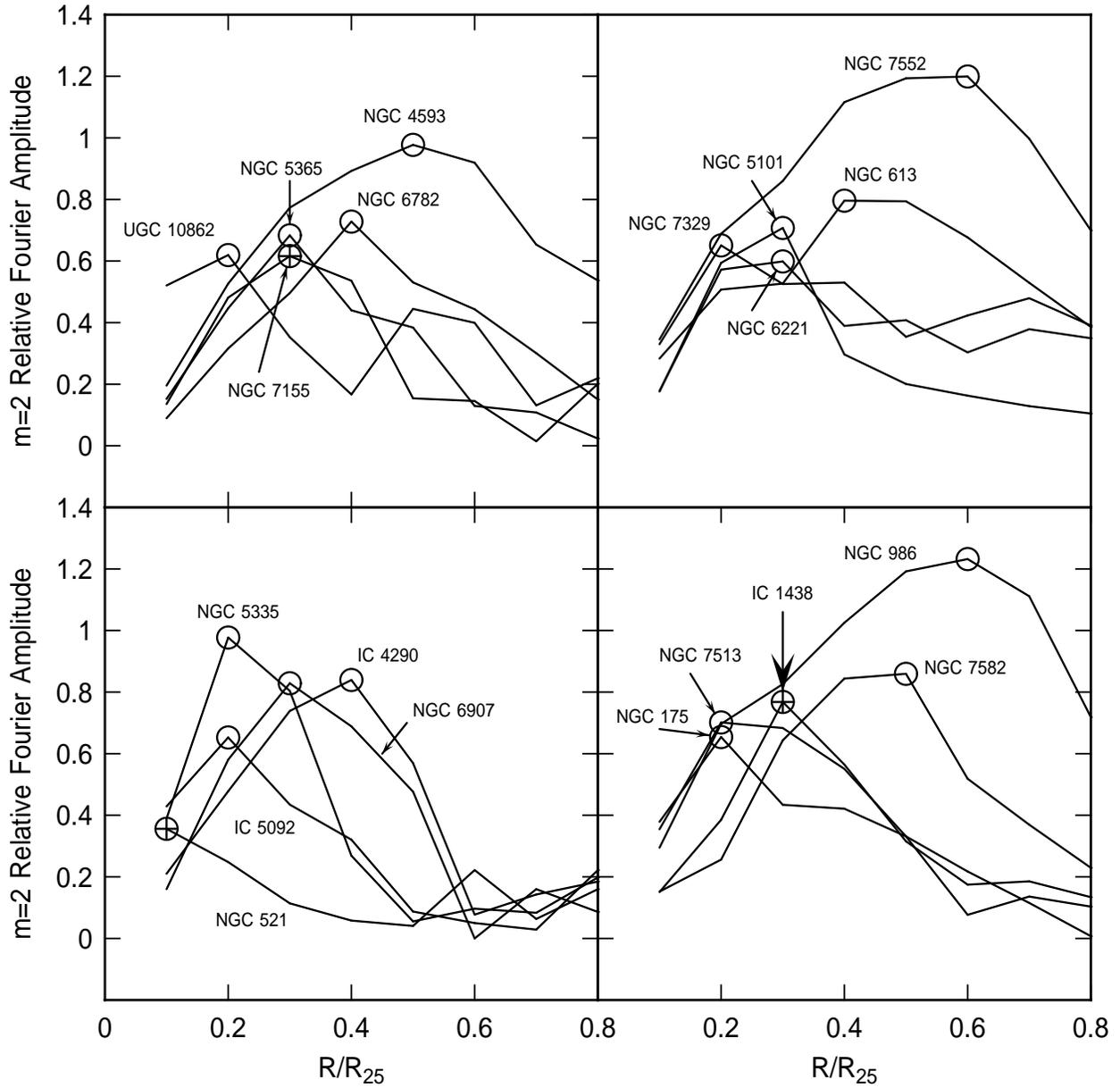}\caption{Relative amplitude of the $m=2$ Fourier
component of the bar and spiral pattern versus radius for 20
galaxies. Circles show the peaks. Circles with plus signs are bars
with exponential intensity profiles, the others have relatively flat
profiles.}\label{fig:20gal}
\end{figure}

Bar and spiral arm amplitudes were measured from the $m=2$ Fourier
components of azimuthal intensity profiles taken at various radii
from polar plots using the deprojected, star-cleaned,
background-subtracted images (as in Regan \& Elmegreen 1997 and
Block et al. 2004). The $m$=2 Fourier intensity amplitude, $I_2$,
was normalized to the average intensity, $I_0$, at each radius;
$I_2$ is defined to be the amplitude of the sinusoidal fit to the
azimuthal profile. Figure \ref{fig:20gal} shows this normalized
amplitude, $A_2=I_2/I_0$, versus the radius normalized to the
standard isophotal radius $R_{25}$ for each galaxy ($R_{25}$ is half
the diameter $D_{25}$ of the $\mu_B=25$ mag arcsec$^{-2}$ isophote
given by de Vaucouleurs et al. 1991). The 20 profiles have been
divided into four panels for clarity. Figure \ref{fig:20gal} shows
that $A_2$ increases with radius and then decreases. The maximum,
$A_2^{max}$, occurs at a radius which we denote by $R_2$. This
radius is approximately equal to the bar length determined by eye in
all cases.  Theory suggests the two lengths should scale together,
with $R_2$ slightly less than the visible bar length (Athanassoula
\& Misiriotis 2002). A correlation may be seen in Figure
\ref{fig:20gal} in the sense that galaxies with higher $A_2^{max}$
also have larger radii at this peak (the peaks are indicated by the
circles; empty circles are flat bars and circles with plus-signs are
exponential bars).

\begin{figure}\epsscale{0.7}
\plotone{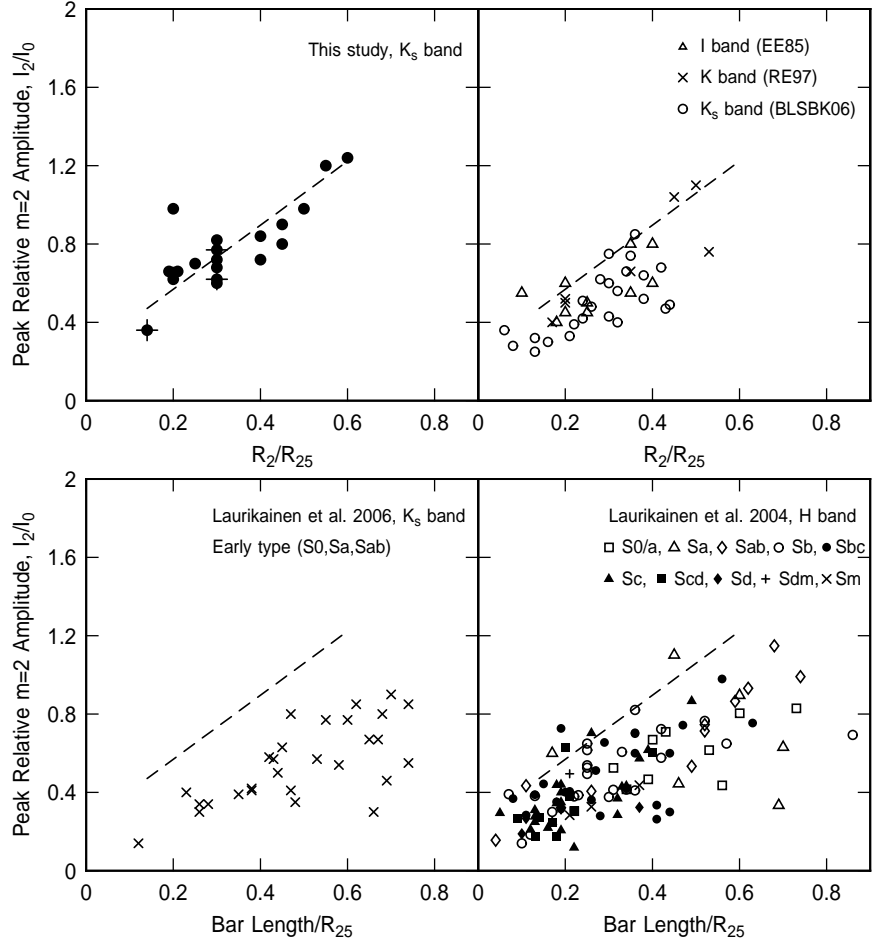}\caption{Peak relative amplitude of the $m=2$
Fourier component versus the normalized radius at the peak. Left:
galaxies in Fig. 1 all imaged in $K_s$ band (plus signs are
exponential bars). Slight displacements among a triplet at
$R_2/R_{25}=0.2$ are for clarity. Published measurements from other
surveys are shown in the top right and bottom. Top right: three
surveys in various passbands having little overlap with galaxies in
the present study (see text). Bottom left: 28 early type (S0,Sa,
Sab) galaxies imaged in $K_s$ band in a combined southern and
northern survey by Laurikainen et al. (2006). Bottom right: 104
galaxies imaged in H-band with the Ohio State Bright Galaxy Survey
and 9 galaxies imaged in H-band with 2MASS, all of various Hubble
types, using measurements in Laurikainen et al. (2004).  All surveys
indicate that bars that are longer compared to their galaxy size
have higher peak relative $m=2$ Fourier amplitudes. This correlation
is present even for early type galaxies. A second, well-known
correlation between bar length and Hubble type is evident from the
lower right panel where the late Hubble types (solid symbols, plus
and cross) tend to have shorter and weaker bars than the early types
(open symbols).}\label{fig:correl}
\end{figure}

This correlation is shown in Figure \ref{fig:correl} (top left),
which plots $A_2^{max}$ versus the normalized radius $R_2/R_{25}$.
The dashed line is a bi-variate least squares fit, repeated in the
other panels.  Longer bars are higher amplitude in relative
intensity. This is sensible considering the general exponential
decline of disk intensity: longer bars extend further out in the
disk, placing their ends where the average background is fainter.
For example, each radial interval of $\sim0.25R/R_{25}$ corresponds
to about one exponential scale length in most galaxies, which is a
factor of 2.7 in disk brightness. This factor is only slightly
larger than the increase in Figure \ref{fig:correl}.  Thus, growing
bars can stay somewhat flat in their intensity profile and still
increase their relative amplitude along with their length because
the surrounding disk is decreasing with radius. Bars apparently grow
relative to the disk size even if the disk grows too because of
angular momentum transfer from the bar (Valenzuela \& Klypin 2003).

Figure \ref{fig:correl} (top right) includes three previous surveys
in which this correlation was present but not noticed. The crosses
are from $K$-band images of 8 different barred galaxies studied by
Regan \& Elmegreen (1997), the circles are from $K_s$-band images of
24 different early type (S0-Sa) barred galaxies in Buta et al.
(2006), and the triangles are from 10 $I$-band images of different
galaxies in Elmegreen \& Elmegreen (1985). Among these three
samples, there are only 3 overlapping galaxies and they are only
between the 1985 and 1997 surveys. The Regan \& Elmegreen
$A_2^{max}$ values were multiplied by 2 because they used the
standard definition of a Fourier component, which, for example,
gives a relative value of 0.5 for an azimuthal profile of
$1+\sin(2\theta)$. We and the other references in Figure
\ref{fig:correl} use twice the Fourier component to reflect the
amplitude of the sinusoidal part of the profile.

The lower panels of Figure \ref{fig:correl} show correlations
present in data from two other studies of bar Fourier amplitudes.
The lower left panel shows data from Laurikainen et al. (2006), who
determined the Fourier amplitudes and bar radii for 28 early type
galaxies (S0,Sa, Sab) in $K_s$ band. The lower right panel shows
data from Laurikainen et al. (2004), who used the Ohio State Bright
Galaxy Survey and 2MASS to measure the H-band properties of 113
galaxies of various Hubble types. Their tabulations give the bar
lengths, not the radii at the peak of the Fourier amplitude. Bar
length is slightly larger than $R_2$, so the points are shifted to
the right of the dashed lines in the figures. Also, $A_2$ is lower
for S0 galaxies than other early types, which lowers some of the
points in the lower left panel (Laurikainen, Salo, \& Buta 2004).
The present correlation was not noticed in either study but it is
present in the data.

Our previous study of $K_s$-band images for 17 barred galaxies
(Block et al. 2004) found a length-amplitude correlation related to
the present one. There we plotted the bar/interbar intensity
contrast at 0.7 bar length versus the deprojected length of the bar
(determined by eye). There was no overlap in galaxies with the
present or the Buta et al. (2006) samples, and only one overlap each
with the Regan \& Elmegreen (1997) and Elmegreen \& Elmegreen (1985)
samples. The bar/interbar intensity contrast was shown by Block et
al. to correlate with the relative amplitude of the $m=2$ Fourier
component, and with the bar torque parameter, $Q_b$. This previous
study discussed the length-amplitude correlation in a different
context, however, noting that the long and high-amplitude bars
tended to be early Hubble type and flat-profile, while the short and
low-amplitude bars tended to be late Hubble type and exponential.
This is true in general, but the present result is in addition to
that. In the present work, the length-amplitude correlation is
present even for the flat bars, and there is no strong correlation
with Hubble type because most of our galaxies are flat-barred.

\begin{figure}\epsscale{1}
\plotone{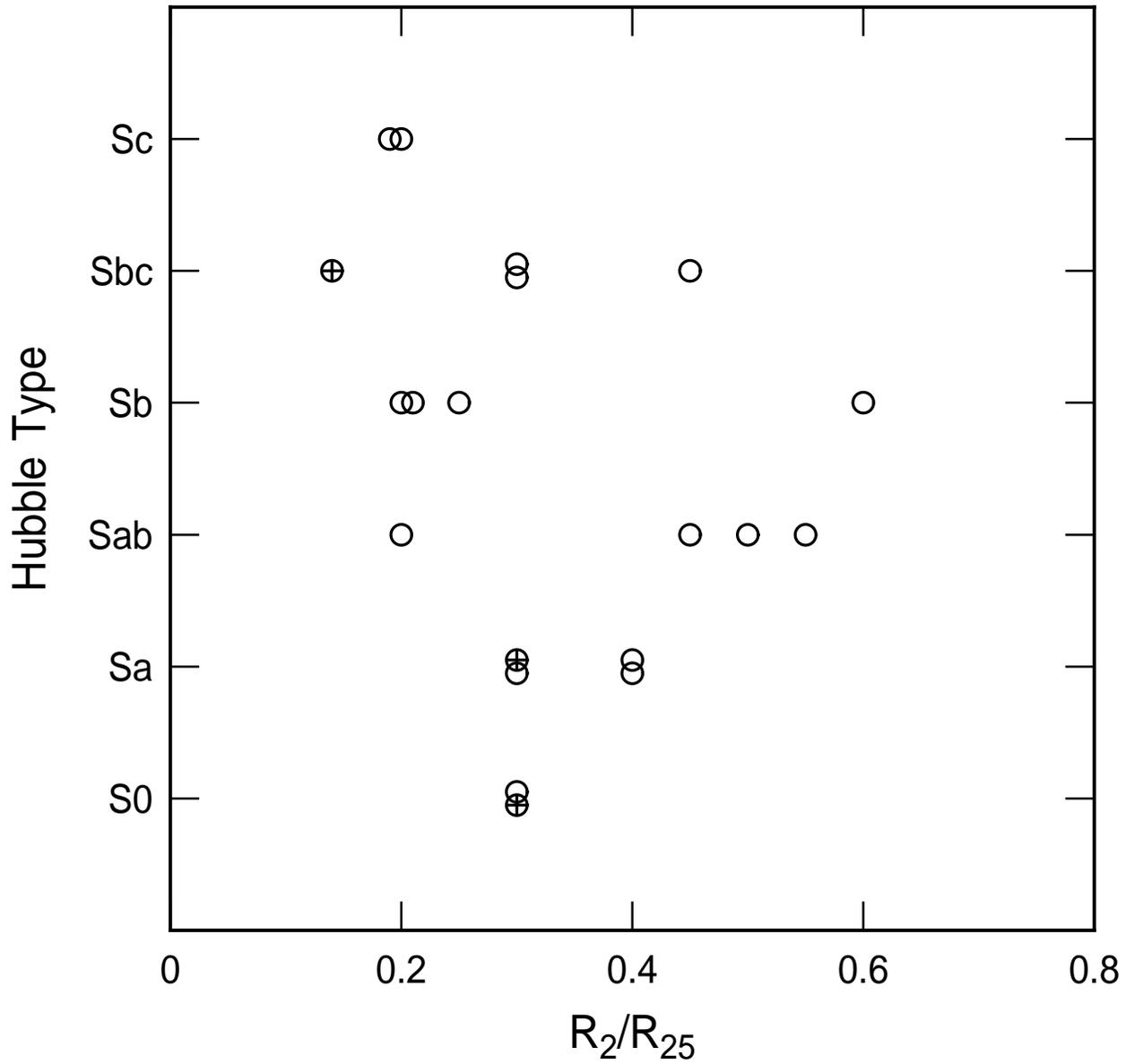}\caption{Hubble type versus the normalized radius at
the peak relative $m=2$ Fourier component (i.e., the relative bar
size) showing little correlation in our sample. The circles with
plus signs are galaxies with exponential bars; the rest have
flat-profile bars. Slight displacements are for
clarity.}\label{fig:ht}
\end{figure}

The $R_2/R_{25}$ length is plotted versus Hubble type for our sample
in Figure \ref{fig:ht}. The circled plus-signs are exponential bars,
and the rest are flat bars. Most of the galaxies in our current
sample are Hubble types Sbc or earlier. The three exponential bars
in our sample have slightly weaker Fourier components than the
average for the flat bars (Fig. \ref{fig:correl}). Evidently, there
are two length-amplitude correlations: one discussed by Block et al.
differentiating early and late type bars (which is presumably
related to different bar resonances; Combes \& Elmegreen 1993), and
another found here that remains even for early-type, flat bars. The
lower right panel of Figure \ref{fig:correl} illustrates these two
correlations in another way by plotting the various Hubble types
with different symbols. The late types tend to be confined to the
lower left in the figure, while the early types display the full
range of bar lengths and amplitudes.

Laurikainen, Salo, \& Buta (2004) found no correlation between the
peak relative torque, $Q_g$, normalized to the radial force, and the
relative radius at the peak of this torque. The relative torque is a
combination of the azimuthal bar amplitude, which determines the
torque, and the radial force from the bulge, which is used to
normalize this torque. Stronger-bulge galaxies have weaker bar
torques for the same relative $m=2$ component. Bulges do not affect
the peak $A_2$ much because the bulge intensity at the end of the
bar is small. On the other hand, bulges do affect $Q_g$ because the
radial force from the bulge is still large at the bar end.

\begin{figure}\epsscale{1}
\plotone{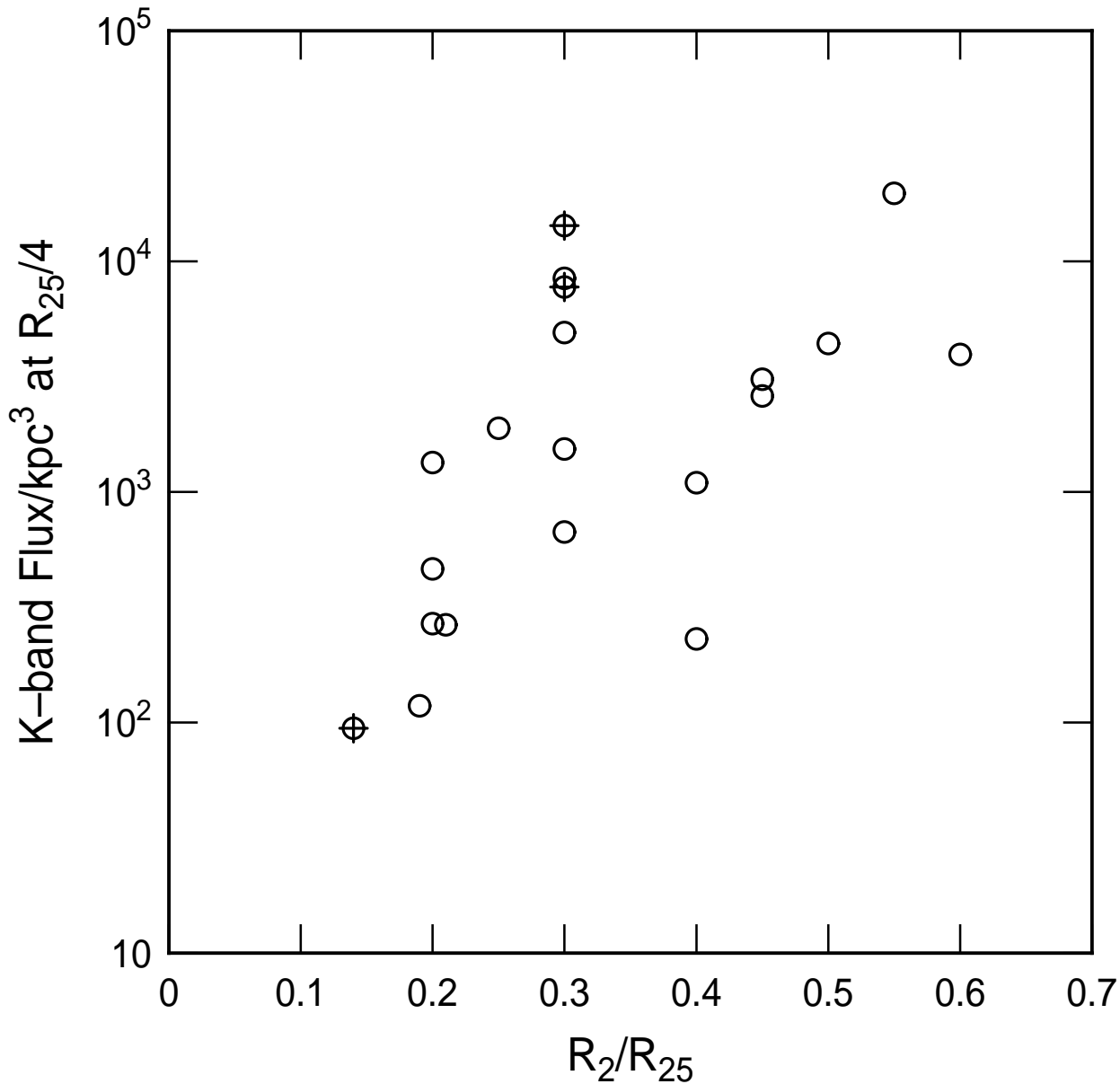}\caption{Luminosity density at $R_{25}/4$ versus the
normalized radius at the peak relative $m=2$ component. Luminosities
are in units of the count rate per unit solid angle integrated
inside $R_{25}/4$, and are calibrated to the same intensity scale
for all galaxies. Plus signs indicate exponential bars. There is a
correlation in which relatively longer bars have denser galactic
centers, suggesting bar growth during a Hubble
time.}\label{fig:density}
\end{figure}

The central luminosity densities of the galaxies were measured from
the $K_s$-band luminosities inside $0.125R_{25}$, $0.25R_{25}$, and
$0.5R_{25}$. The $K_s$-band is dominated by old stars and traces the
mass fairly well if dark matter is not significant there. Most of
the galaxies are early type and centrally condensed so the 3
luminosities measured in this way were all about equal. Because the
$R_2/R_{25}$ lengths vary from $\sim0.1R_{25}$ to $\sim0.5R_{25}$,
and we want a representative density in the bar region, we use the
luminosity inside $0.25R_{25}$. The central density is then taken to
be this luminosity divided by the cube of the radius at
$0.25R_{25}$, measured in kpc using the distances in Table 1 (from
the galactocentric GSR in the NASA/IPAC Extragalactic Database).
Figure \ref{fig:density} shows the central $K_s$-band density versus
the normalized radius at the peak $m=2$ amplitude (plus signs denote
exponential bars). There is a correlation in the sense that longer
bars occur in denser galaxies.

These two correlations provide new information to supplement
properties found in other bar correlations. Athanassoula \& Martinet
(1980) and Martin (1995) found a correlation between the lengths of
bars and bulges, and Elmegreen \& Elmegreen (1985) found a
correlation between bar length, amplitude, and early versus late
Hubble types, as mentioned above (see review in Ohta 1996).

\section{Discussion}

We find that among fairly early type galaxies, relative bar length
and relative $m=2$ intensity correlate with each other but not
obviously with the Hubble subtype. The lengths and amplitudes also
correlate with the central luminosity density of the galaxy. These
correlations are in the sense expected by numerical simulations
which suggest that angular momentum gradually transfers from a bar
to the surrounding disk, bulge, and halo (see Athanassoula 2003 and
references therein).  With a loss of angular momentum, bars should
slow down, and this means their corotation radii move outward. The
stellar orbits in the bar should also get more elongated as angular
momentum is proportional to the orbital area, and this translates to
ellipticity for a constant orbital energy. As the orbital
ellipticity increases, the stars become more concentrated in the bar
and the bar gets stronger. If the orbits also scatter in energy,
then their semi-major axes should grow too, following the moving
corotation resonance. In this case, bars would grow in length as
they get higher relative amplitudes during angular momentum loss.
This is apparently what we observe here.

The correlation with central density is consistent with angular
momentum loss because galaxies with higher central densities evolve
faster. In a given galaxy lifetime, the bars which evolve faster
will have transferred more of their angular momentum outward and at
the present time will have longer and higher-amplitude bars. The
correlation with central density could also result from a larger
reservoir for bar angular momentum in the larger bulges. An inverse
process might be responsible too, where a strong bar forms first and
this causes the bulge to grow through accretion (e.g., Athanassoula
1992, 2003).

The lack of a correlation between relative bar length and peak
relative bar torque $Q_g$ may be understood from our correlations
with central density.  For a given bulge, angular momentum transfer
should increase both the peak amplitude and the peak torque of the
bar over time. Galaxies with denser bulges do this faster, so at any
given time, the peak amplitude correlates with bulge density.
However, denser bulges weaken $Q_g$ because this quantity is
normalized to the radial force (Laurikainen, Salo, \& Buta 2004).
This normalization offsets the increasing bar amplitude that comes
from angular momentum transfer. As a result, $Q_g$ does not show the
same correlations as the $m=2$ Fourier amplitude.

Galaxies with dense bulges should not have bars if bulges prevent
bar formation or growth (e.g., Sellwood 1980).   However, our data
show that high central densities correlate with high-amplitude bars.
The observed correlation suggests that bars and bulges grow
together, in agreement with Sheth et al. (2007).

\section{Conclusions}

Bars in intermediate and early type spirals have a correlation
between their relative lengths and their relative $m=2$ Fourier
components, and both increase with the central density. These
correlations are consistent with models in which bars lose angular
momentum to the surrounding disk, bulge, and halo over long periods
of secular evolution. The bars contain very old stars and must have
been present for a high fraction of the Hubble time, like the
bulges.

\acknowledgements We thank Emma Allard for help during the
observations and with the data reduction, and Stuart Ryder for
excellent support at the AAT. We thank Heikki Salo and Eija
Laurikainen for useful comments on the manuscript. Helpful comments
by the referee are appreciated.  DME thanks Vassar College for
publication support through a Research Grant. RB acknowledges the
support of NSF grant AST 05-07140. I.P. acknowledges support from
the Mexican foundation CONACyT under project 35947­.E.   This
research has made use of the NASA/IPAC Extragalactic Database (NED)
which is operated by the Jet Propulsion Laboratory, California
Institute of Technology, under contract with the National
Aeronautics and Space Administration.

\clearpage

\begin{deluxetable}{lccccccc}
\tabletypesize{\scriptsize} \tablewidth{0pt}
\tablecaption{Barred Galaxies\label{table1}}
\tablehead{Galaxy&typetablenotemark{a}&D (Mpc)&R$_{25}$ (arcsec)&$R_2/R_{25}$\tablenotemark{b}&$I_2/I_0$\tablenotemark{c}\\
&&&&&&} \startdata
NGC175   &SB(\b{r}s)ab&53.9&64.1&0.2&0.33\\
NGC521   &SB(\b{r}s)bc&   69.6    &   94.9    &  0.15    &   0.18\\
NGC613   &SB(rs)bc&   19.8    &   164.9   &  0.5     &   0.4\\
NGC986   &(R$^\prime_1$)SB(rs)b&   25.7    &   116.7   &  0.6     &   0.62 \\
NGC4593  &(R$^\prime_1$)SB(rs)ab&   35.6    &   116.7   &  0.5     &   0.48\\
NGC5101  &(R$_1$R$^\prime_2$)SB(\b{r}s)a&   23.7    &   161.1   &  0.3     &   0.36\\
NGC5335  &SB(r)b&   63.2    &   64.1    &  0.2     &   0.5\\
NGC5365  &(R)SB0$^-$&   31.6    &   88.5    &  0.3     &   0.34\\
NGC6221  &SB(s)bc pec&   19      &   106.4   &  0.3     &   0.3\\
NGC6782  &(R$_1$R$^\prime_2$)SB(r)a&   52.6    &   65.6    &  0.4     &   0.37\\
NGC6907  &SAB(s)bc&   44.5    &   99.3    &  0.3     &   0.42\\
NGC7155  &SB(r)0$^o$&   26.7    &   65.6    &  0.3     &   0.31\\
NGC7329  &SB(r)b&   43.1    &   116.7   &  0.2     &   0.33  &\\
NGC7513  &SB(s)b&   21.9    &   94.9    &  0.25    &   0.35  &\\
NGC7552  &(R$^\prime_1$)SB(s)ab&   21.7    &   101.7   &  0.55    &   0.6\\
NGC7582  &(R$^\prime_1$)SB(s)ab&   21.3    &   150.4   &  0.45    &   0.45\\
IC1438   &(R$_1$R$^\prime_2$)SAB(r)a&   20      &   72      &  0.3     &   0.385\\
IC4290   &(R$^\prime$)SB(r)a&   64.3    &   47.6    &  0.4     &   0.42\\
IC5092   &(R)SB(s)c&   43.3    &   86.5    &  0.2     &   0.33\\
UGC10862 &SB(rs)c&   24.8    &   82.6    &  0.2     &   0.31\\
\enddata
\tablenotetext{a}{Classifications are either from the de Vaucouleurs
Atlas of Galaxies (Buta, Corwin, \& Odewahn 2007) or estimated by RB
in the same system based on available image
material.}\tablenotetext{b}{Relative radius of peak relative $m=2$
Fourier amplitude.} \tablenotetext{c}{Peak relative $m=2$ Fourier
amplitude.}
\end{deluxetable}

\clearpage

\clearpage

\clearpage

\clearpage

\end{document}